\documentclass[fleqn,twoside,twocolumn,nofootinbib,showkeys]{revtex4} 
\usepackage[sec,nocpr]{ujp} 

\begin{document}
\title[Magnetic Field of Cosmic Strings]
{MAGNETIC FIELD OF COSMIC STRINGS\\ IN THE EARLY UNIVERSE}%
\author{L.V. Zadorozhna}
\affiliation{Taras Shevchenko National University of Kyiv, Faculty of Physics}
\address{2, Academician Glushkov Ave., Bld.~1, Kyiv 03127, Ukraine}
\email{Zadorozhna_Lida@ukr.net}
\author{B.I. Hnatyk}
\affiliation{Taras Shevchenko National University of Kyiv, Astronomical Observatory}
\address{3, Observatorna Str., Kyiv 04053, Ukraine}
\email{hnatyk@observ.univ.kiev.ua}
\author{Yu.A.~Sitenko}%
\affiliation{Bogolyubov Institute for Theoretical Physics, Nat.
Acad. of
Sci. of Ukraine}%
\address{14b, Metrolohichna Str., Kyiv 03680, Ukraine}%
\email{yusitenko@bitp.kiev.ua}

\udk{524.7} \pacs{98.80.Cq} \razd{\secxi}

\autorcol{L.V.\hspace*{0.7mm}Zadorozhna, B.I.\hspace*{0.7mm}Hnatyk,
Yu.A.\hspace*{0.7mm}Sitenko}

\setcounter{page}{398}%

\begin{abstract}
Cosmic strings are topological defects which can be formed as a
result of phase transitions with a spontaneous symmetry breaking in
the early Universe. The possibility of the generation of a magnetic
field around a cosmic string on the Grand Unification energy scale
(GUT scale) in the early Universe immediately after the termination
of the deconfinement-confinement phase transition has been analyzed.
It is found that a circular current and a magnetic field directed
along the string are induced around the string in the vacuum of a
pseudoscalar matter consisting of charged pions. We also have
studied the interaction between the magnetic flux tube surrounding
the string (the string magnetosphere) and the cosmic plasma in the
early Universe. A possibility of magnetization of the cosmic plasma
surrounding the string owing to its interaction with the string
magnetic field has been analyzed.
\end{abstract}

\keywords{cosmic string, phase transitions, vacuum polarization
effect, ultrarelativistic plasma, vacuum of pseudoscalar matter
consisting of charged pions, bow shock, magnetic~tube.}

\maketitle

\section{Introduction}

According to the standard cosmological model, the Universe is
expanding and cooling down since the Big-Bang moment, but remains,
as a whole, uniform and isotropic. There are the reasons to consider
the Universe to pass through a chain of phase transitions in the
course of its cooling \cite{kampfer, rocher}. The phase transition
associated with the separation of the strong interaction from the
electroweak one~-- the end of Grand Unification Epoch~-- occurred in
10$^{-35}~\mathrm{s}$ after the Big-Bang time moment, at a
temperature of $2\times10^{16}~\mathrm{GeV}$. This phase transition
was accompanied by the symmetry breaking: from a higher one
characteristic of the unified interaction to symmetries inherent to
plasma components at low temperatures \cite{wein}. The expansion of
new phase regions, which initially had no causal links with various
vacuum states stemming from the spontaneous symmetry breaking, can
give rise to the emergence of topological defects at the interfaces
between those regions. Cosmic strings are a type of topological
defects, which can be formed owing to the phase transition with a
spontaneous symmetry breakdown in the early
Universe~{\mbox{\cite{kibb, vilenk,
vilenpp}}.

In 10$^{-10}~\mathrm{s}$ after the Big Bang, the electroweak
interaction separated. When the temperature fell down below 124~GeV,
there emerged a phase with violated electroweak symmetry with the
nonzero Higgs condensate and massive $W^{\pm}$ and $Z$ bosons. The
deconfinement--confinement transition, i.e. from a quark-gluon
plasma to hadrons, took place in 10$^{-5}~\mathrm{s}$ after the
Big-Bang time moment, when the matter got cooled down to a
temperature below \mbox{200$\mathrm{~MeV}$~\cite{simonschettler,
pauchyhwang}.}

Linear defects~-- cosmic strings~-- are formed in the overwhelming
majority of theoretical models dealing with the early Universe
\cite{rocher}. The topologically stable strings have no ends, i.e.
they can be infinite or form closed loops. The specific mass per
unit string length and the string tension (hereafter, we used the
fundamental unit system, where $\hbar=c=k_{\rm B}=1$) are of the
order of $\mu\sim\eta^{2}$, where $\eta$ is the energy scale of the
symmetry breakdown. It is determined by the phase transition
temperature in the Universe and, in turn, determines the Higgs field
mass, $m_{\rm H}\sim\eta$.

For strings of the Grand Unification scale, the specific mass per unit length
amounts to 10$^{22}~\mathrm{g/cm}$. The transverse string radius $r_{0}$ is
determined from the relation $r_{0}m_{\rm H}\sim1,$ and $r_{0}%
\sim10^{-30}$~$\mathrm{cm}$ for GUT-strings. As a result of the substantial tension, the
string segments move at the velocities $V_{s}=\beta_{\rm s}c$ close
to the velocity of light $c$. The average velocity within the
correlation length approximately amounts to $\left\langle
V\right\rangle \sim0.15c$, and the root-mean-square velocity of
a string in the radiation-dominating epoch is
$V_{\mathrm{rms}}\sim0.62c$ \cite{vilenk}.

\section{Magnetic Field around a GUT-String}

Provided a charged field inside a string, the latter can behave as a
superconductor \cite{ost}. Therefore, when a superconducting string moves, for
example, in the intergalactic magnetic field, a current is generated that
flows along the string. As a result, there arises the magnetosphere around the
string; it is the own magnetic field of the string. If such a superconducting
string moves through a cosmic plasma at relativistic velocities, the
interaction of its magnetic field with the cosmic plasma will stimulate the
generation of a shock wave around the string \cite{chudetal86}. At the shock
wave front, particles of the cosmic plasma will be accelerated to high
energies. As a result, they will emit electromagnetic waves in a wide range of
energies. In works \cite{zadoree, zadoreb}, the generation of non-thermal
radiation at the interaction between a superconducting string and the cosmic plasma
was described in detail.

In work \cite{sitenko} (see also works \cite{sitcab, sitbozvac}), it
was shown that a magnetic field can be generated even near the
surface of an ordinary non-superconducting string due to vacuum
polarization effects in the quantized field of a charged matter
around the string. Namely, the local cosmic string characterized by
the tension $\mu\sim m_{\rm H}^{2}$ and the flux $\Phi$ of the gauge
field in the string induces a current $j$ that circulates in vacuum
around the string and a magnetic field $B$ directed along the
string. The both quantities fall down exponentially at large
distances from the string, being connected by the relation
\begin{equation}
B(r)=\int\limits_{r}^{\infty}dr\frac{\nu}{r}ej(r),
\end{equation}
where $\nu=(1-4G\mu)^{-1}$, and $e$ is the electric charge of the quantized matter
field. In the case of a field with mass $m$ and zero spin, the total
flux of the induced magnetic field equals
\begin{equation}
\Phi_{\mathrm{B}}=\frac{{e}}{6\pi}\left(\!F-\frac
12\!\right)F(1-F)\nu^2\ln \frac{m_{\mathrm{H}}}{m},
\end{equation}
where $F=e\Phi\left(2\pi \hbar c\right)^{-1} -[\![{e}\Phi(2\pi \hbar
c)^{-1}]\!]$, and $\left[\!\left[u\right]\!\right]$ denotes the
integer part of $u$ \cite{sitenko}.

Consider a cosmic string of the Grand Unification scale (a
GUT-string) and its influence on the pseudoscalar matter vacuum
consisting of charged pions. Such matter arises at an early stage of
Universe's evolution, right after the phase transition
``deconfinement--confinement'' has terminated, as a result of the
binding of the quarks $u$ ($\bar{u}$) and $\bar{d}$ ($d$) into
$\pi^{\pm}$-mesons \cite{nakamura}. A magnetic field is induced in
vacuum around the string. According to the results of work
\cite{sitenko}, this field is given by the expression
\[
B(r)\approx\frac{{e}\left[F\sin((1-F)\pi)
-(1-F)\sin(F\pi)\right]}{2(4\pi)^{2}}\frac{\hbar
c}{E_{\pi^{\pm}}}\times
\]\vspace*{-5mm}
\begin{equation}
 \times\frac{{e}^{-2\frac{E_{\pi^{\pm}}}{\hbar c}r}}{r^3},
\end{equation}
where $E_{\pi^{\pm}}=m_{\pi^{\pm}}c^{2}$, and $m_{\pi^{\pm}}$ is the
mass of
the charged pionic field. We also took into account that $\nu_{\mathrm{GUT}%
}\approx1$. For $F$-values corresponding to the maximum value of
$\Phi_{\rm B}$ (in particular, $F_{1}\approx0.8$ and
$F_{2}\approx0.2$), we obtain
\begin{equation}
B(r)=B_{0}\frac{{e}^{-2r/r_{\rm B}}}{\left(  r/r_{\rm B}\right)  ^{3}}%
\approx2.7\times10^{13}\frac{{e}^{-2r/r_{\rm B}}}{\left( r/r_{\rm
B}\right)
^{3}}~\mbox{(Gs)},\label{mag_f}%
\end{equation}
where we took into consideration that a characteristic scaling
factor $r_{\rm B}=\hbar
c/E_{\pi^{\pm}}=1.4\times10^{-13}$~$\mathrm{cm}$ can be introduced
for the magnetic field.

\section{Plasma Parameters around~a~GUT-String}

At the examined early stage of Universe's evolution, the magnetic
field of a string interacts with the surrounding cosmic plasma. At
high temperatures typical of the early Universe, the rate of
reactions between particles exceeds a characteristic time connected
with the rate of Universe's expansion, so that the cosmic plasma is
in thermal equilibrium with electromagnetic radiation. The main
contribution to the Universe energy density is made by
ultrarelativistic particles, for which, under the available
conditions, $m_{i}c^{2}\ll k_{\rm B}T(t)$ and $\mu_{i}=0$, where
$T(t)$ is the equilibrium temperature at the cosmological time
moment $t$, $m_{i}$ is the mass of particles of the $i$-th type, and
$\mu_{i}$ is the corresponding chemical potentials \cite{gorbrub}.
Then, the energy density can be approximated as follows \cite{nakamura}:%
\[
{e}_{\mathrm{th}}=\left(\!\sum\limits_{\mathrm{b}}N_{\mathrm{b}}+\frac{7}{8}
\sum\limits_{\mathrm{f}}N_{\mathrm{f}}\!\right)\frac{\pi^2}{30}\frac{k_{\mathrm{B}}^4T^4}
{\hbar^3c^3}=
\]\vspace*{-5mm}
\begin{equation}
=\frac{\pi^2}{30}N(T)\frac{k_{\mathrm{B}}^4T^4}{\hbar^3c^3},
\end{equation}
where $N_{b}$ and $N_{f}$ are the helicity numbers (the number of
spin projections onto the momentum direction) for every boson and
fermion, respectively; the summation is carried out over all bosonic and
fermionic states; and $N(T)$ is the effective number of degrees of
freedom. The number of degrees of freedom depends on the Universe
composition and, therefore, depends
on the temperature. In particular, $N\left(  m_{\pi^{\pm}}c^{2}<k_{\rm B}%
T<k_{\rm B}T_{\rm c}\right)  =69/4$, where $k_{\rm B}T_{\rm c}=$
$=200\mathrm{~MeV}$ is the temperature of the phase transition
``deconfinement--confinement''. Notice that the energy density $e_{\rm
th}=1.9\times10^{36}~\mathrm{erg/cm}^{3}$ at this temperature.

The relation between the temperature $T$ in the Universe and the time $t$
reckoned from the Big Bang moment during the radiation-dominating epoch is
expressed by the simple formula $tT_{\rm MeV}^{2}=$\linebreak $=2.4\left[  N(T)\right]  ^{-1/2}%
$, where $t$ is measured in second units and $T_{\mathrm{MeV}}$ in
megaelectronvolts \cite{nakamura}. The time corresponding to the
deconfinement--confinement transition equals $t_{\rm
c}=1.4\times10^{-5}~\mathrm{s}$.

From the expression for the concentration of ultrarelativistic particles
\begin{equation}
\label{konz}
n=\left(\!\sum\limits_{\mathrm{b}}N_{\mathrm{b}}+\frac{3}{4}\sum\limits_{\mathrm{f}}N_{\mathrm{f}}\!\right)
\frac{\zeta(3)}{\pi^2}\frac{k_{\mathrm{B}}^3T^3}{\hbar^3c^3},
\end{equation}
where the value of zeta-function $\zeta(3)=1.2$, it is possible to
estimate
the concentration of charged particles (at the temperature $m_{\pi^{\pm}}%
c^{2}<k_{\rm B}T<k_{\rm B}T_{\rm c}$, these particles are $e^{\pm}$,
$\mu^{\pm}$, and $\pi^{\pm}$), $n_{\rm
ch}=1.0\times10^{39}$~\textrm{cm}$^{-3}$, and the average distance
between them $d_{\rm ch}=n_{\rm ch}^{-1/3}$ (at $T=T_{\rm c}$, we
obtain $d_{\rm ch}=9.9\times 10^{-14}~\mathrm{cm})$. The
characteristic scale of a magnetic field, $r_{\rm B}$, turns out
close to the average distance between charged particles, $r_{\rm
B}\sim d_{\rm ch}$.

Let us also estimate the mean free path of a charged particle at
this temperature. The energy of an ultrarelativistic particle is
given by the relation $E\approx cp=$ $=3k_{\rm B}T$, where $p$ is
particle's momentum. The mean free path is
$\lambda\sim1/(\sum_{i}\sigma_{i}n_{i})$, where $n_{i}$ is the
concentration of particles of the $i$-th kind, and $\sigma_{i}$ is
the scattering cross-section for them. Taking into account that all
scattering cross-sections for the electromagnetic interaction
$\sigma_{i}\sim e^{4}/E^{2}$, where $e$ is the electron charge, the
mean free path is of the order $\lambda\sim10^{-8}$~$\mathrm{cm}$
at~$T=T_{\rm c}$.

At $T\sim100\mathrm{~MeV}$ in the cosmic plasma, the concentration of
non-relativistic particles is low in comparison with that of photons. Baryons
are non-relativistic particles, and their total concentration $n_{\rm bar}%
\sim10^{-9}n_{\gamma}\sim10^{29}$~\textrm{cm}$^{-3}$, where $n_{\gamma}$ is
the concentration of photons \cite{gorbrub}. This means that the order of
magnitude for the ratio between the numbers of baryons and photons coincides
with the value of baryon asymmetry in the Universe.

\section{String Magnetosphere and Magnetic Field~Transfer to Cosmic Plasma}

Being surrounded with a shell-like magnetic field (a magnetic flux
tube), the string moves at a typical relativistic velocity through
the cosmic plasma of density $\rho$ (in general, this velocity is
higher than the sound speed in the plasma of ultrarelativistic
particles, $a_{\rm s}=c/\sqrt{3}$). Therefore, in the framework of
the hydrodynamic approximation, the flow of a relativistic (in the
string reference frame) plasma  around the string is similar to a
non-relativistic case of the supersonic solar wind flowing around
Earth's magnetosphere. As a result, there emerges a shock wave in
the incident plasma flow,  a contact discontinuity between the
plasma behind the shock wave and the string magnetosphere
\cite{chudetal86, zadoree}. The shock wave radius along the string
motion direction can be determined from the equality between the
pressure of the incident plasma $P=P_{\rm th}+P_{\rm dyn}$ (here,
$P_{\rm th}$ is the thermal pressure of ultrarelativistic gas,
$P_{\rm dyn}=\gamma_{\rm s}^{2}\rho c^{2}$ is the dynamical
pressure, and $\gamma_{\rm s}$ is the Lorentz factor of the string
and the magnetosphere with respect to the plasma, $\gamma_{\rm
s}^{2}\geq1.5$) and the magnetic field pressure $P_{\rm
B}=B^{2}\left( R_{\rm sh}\right) /8\pi\approx B_{0}^{2}r_{\rm
B}^{6}/(8\pi R_{\rm sh}^{6})$\ in the magnetosphere deformed by the
plasma flux.

In this hydrodynamic scenario of the cosmic plasma flowing around
the string magnetosphere, two channels of magnetic field appearance
in the plasma are possible. One of them is associated with the
emergence of instabilities at the contact discontinuity at the
magnetosphere--downstream flux interface. As a result, some part of
the plasma flux becomes magnetized with a subsequent enhancement of
the field in the turbulent flux to values typical of relativistic
fluxes $e_{\rm B}=\epsilon_{\rm B}e_{\rm th}$; i.e. the energy
density of the turbulent magnetic field ($\epsilon_{\rm
B}\sim0.01\div0.1$) becomes comparable with the density of the
thermal plasma energy $e_{\rm th}$. The other channel is related to
the emergence of a magnetic field in a vicinity of the shock wave in
the non-magnetized plasma. This field will be removed into the
region behind the shock wave, and its magnitude will depend on the
relation between dissipative processes and processes that strengthen
the field.

However, in the situation of the interaction between the string
magnetic field and the cosmic plasma, which is considered here,
the application of the hydrodynamic approximation is problematic.
By equating the pressure of the incident plasma $P=P_{\rm
th}+P_{\rm dyn}$ and the pressure of the magnetic field $P_{\rm
B}\left( R_{\rm sh}\right) $, we can calculate the shock wave
radius in the hydrodynamic
approximation,%
\begin{equation}
\label{radius_sh} R_{\mathrm{sh}}\approx
r_{\mathrm{B}}\left(\!\frac{B_{0}}{\sqrt{8\pi
P}}\!\right)^{\!1/3}=2.4\times10^{-15}~\text{cm}.
\end{equation}
For the hydrodynamic approximation to be valid, the sizes of a shock
wave and a contact discontinuity must substantially exceed the mean
free path of plasma particles. However, the condition of
hydrodynamic approximation eligibility is not satisfied in our case,
and the interaction between the string and plasma particles is
reduced to separate scattering events of charged particles (the
deflection of their trajectories) in the magnetic field of a string.
In this case, the transfer of the magnetic field into the plasma
should be considered separately. At the qualitative level, we may
assume that, when the incident plasma flows around string's magnetic
flux tube, it captures some part of the magnetic flux formed by
field lines located at farther distances from the string than the
average distance between particles $d_{\rm ch}$. Since the Hubble
volume (about $r_{h}^{3}\sim(ct)^{3}$) includes about $r_{h}\sim ct$
of the string length, the plasma drifts a magnetic field with flux
$d\Phi_{\rm cap}/dt\sim V_{\rm s}d_{\rm ch}B(d_{\rm ch})$ and energy
$dW_{\rm cap}/dt=$ $=V_{\rm s}r_{h}d_{\rm ch}( B^{2}(d_{\rm
ch})/8\pi) $ during a time unit. Owing to a rapid decrease of the
magnetic field with increasing the distance from the string, the
effective
time of magnetic field energy transfer into the plasma equals $\Delta t\sim t_{\rm c}%
$, and the transferred energy in the Hubble volume, $\Delta W_{\rm
cap}\sim\left( dW_{\rm cap}/dt\right)  \Delta t\sim\beta_{\rm
s}r_{h}^{2}(t_{\rm c})d_{\rm ch}e_{\rm B}(d_{\rm ch})$, represents a
very insignificant fraction of the thermal energy, $\sim$$\left( d_{\rm
ch}/r_{h}(t_{\rm c})\right)  (e_{\rm B}(d_{\rm ch})/e_{\rm
th}(t_{\rm c}))$\,$\sim$\,$10^{-18} \times$
$\times10^{-11}$\,$\sim$\,$10^{-29}$. The further counteraction
between the dissipation processes and the dynamo-processes that
strengthen the magnetic field will be responsible for the final
contribution of cosmic strings to the generation of the observed
cosmological magnetic \mbox{field \cite{kandus}.}

\begin{figure}
\vskip1mm
\includegraphics[width=\column]{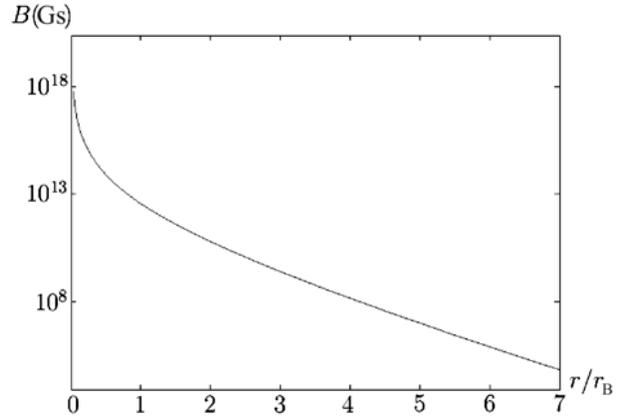}
\vskip-4mm\caption{Magnetic field $B$ at various distances $r$ from
the string  }
\end{figure}

\section{Discussion and Conclusions}

In this work, the generation of a magnetic field around a cosmic
string of {the Grand Unification energy scale} in the early Universe
at the times after the phase transition
``deconfinement--confinement'' is considered. The magnetic field is
induced around a cosmic string in the vacuum of a pseudoscalar
matter consisting of charged pions; this field is directed along the
string. In Figure, the variation of the magnetic field with the
distance from the string is depicted. The interaction between the
magnetic field around the string and the ultrarelativistic cosmic
plasma was studied. In particular, the parameters of the magnetic
field and the thermodynamic characteristics of the plasma in the
early Universe after the phase transition
``deconfinement--confinement'' are determined. The characteristic
scale of the formed magnetic field flux tube is shown to be
comparable with the average distance between charged particles in
the plasma and smaller than their mean free path length. Therefore,
at typical relativistic velocities of the string in the cosmic
plasma, the latter will reduce the magnetic field only at rather
long distances from the string; these are distances of the order of
those between plasma particles, where the magnetic field is already
suppressed substantially. As a result, only small fractions of the
magnetic field flux and energy are transferred into the plasma.
However, the resulting value of the field transferred into plasma
will also depend on the subsequent field evolution governed by both
the dissipative and enhancing (dynamo) \mbox{processes.}\looseness=1

\vspace*{-5mm}
\rezume{%
Л.В. Задорожна, Б.І.~Гнатик, Ю.О.~Ситенко}{МАГНІТНЕ ПОЛЕ КОСМІЧНИХ
СТРУН\\ У РАННЬОМУ ВСЕСВІТІ} {Космічні  струни~-- топологічні
дефекти, що могли утворюватися під час фазових переходів зі
спонтанно порушеною симетрією у ранньому Всесвіті. В роботі
розглянута можливість утворення магнітного поля навколо струни
енергетичного масштабу Великого Об'єднання в умовах раннього
Всесвіту, одразу після фазового переходу деконфайнмент--конфайнмент.
Навколо космічної струни у вакуумі псевдоскалярної матерії, що
складається із заряджених піонів, індукуються коловий струм та
магнітне поле, напрямлене вздовж струни. В роботі досліджено
взаємодію магнітної силової трубки, що оточує струну~-- магнітосфери
струни~-- з космічною плазмою в ранньому Всесвіті. Проаналізовано
можливість замагнічення оточуючої струну плазми внаслідок її
взаємодії з магнітним полем струни.}

\end{document}